\documentclass[12pt]{article}
\usepackage[left=1in,right=1in,top=1in,bottom=1in]{geometry}
\usepackage{graphicx}
\usepackage{float}
\usepackage{cite}
\usepackage{amsfonts}
\usepackage{amssymb}
\usepackage{amsmath}
\usepackage{relsize}
\usepackage[hidelinks]{hyperref}
\usepackage{comment}
\usepackage{soul}
\usepackage{color}
\usepackage{cleveref}
\usepackage{comment}

\def\gtwid{\mathrel{\raise.3ex\hbox{$>$\kern-.75em\lower1ex\hbox{$\sim$}}}}
\def\ltwid{\mathrel{\raise.3ex\hbox{$<$\kern-.75em\lower1ex\hbox{$\sim$}}}}
\def\square{\kern1pt\vbox{\hrule height 1.2pt\hbox{\vrule width 1.2pt\hskip 3pt
   \vbox{\vskip 6pt}\hskip 3pt\vrule width 0.6pt}\hrule height 0.6pt}\kern1pt}
\newcommand{\p}{\partial}
\newcommand{\e}{\varepsilon}

\renewcommand{\e}{\varepsilon}

\crefname{equation}{Eq.}{Eqs.}
\Crefname{equation}{Equation}{Equations}

\begin{document}
\begin{titlepage}

\begin{flushright}
UFIFT-QG-26-05
\end{flushright}

\vskip 3cm

\begin{center}
{\bf Sensing the Inflationary Production of Scalars}
\end{center}

\vskip 2cm

\begin{center}
A. J. Foraci$^{*}$, C. Litos$^{\dagger}$ and R. P. Woodard$^{\ddagger}$
\end{center}

\begin{center}
\it{Department of Physics, University of Florida, \\
Gainesville, FL 32611, UNITED STATES}
\end{center}

\vskip 2cm

\begin{center}
ABSTRACT
\end{center}

We review the mechanism by which loops of matter fields contribute to the 
graviton self-energy during de Sitter inflation. This self-energy is used to
quantum-correct the linearized Einstein equations. A Green's function method 
is employed to obtain exact 1-loop corrections to the plane wave mode 
functions of gravitational radiation, subject to the usual ambiguity in the 
initial state. Conformally coupled matter, which does not experience 
inflationary particle production, makes only a logarithmic enhancement 
of the rate at which the imaginary part of the mode function goes to zero 
after horizon crossing. These corrections can be understood, and even summed 
up, using a variant of the renormalization group. However, massless, 
minimally coupled scalars, which experience massive inflationary particle
production, induce a much stronger enhancement of the rate at which the 
real part of the mode function approaches a constant. One interpretation of
this effect is as a shift of the inflationary Hubble parameter.

\begin{flushleft}
\end{flushleft}

\vskip 3cm

\begin{flushleft}
$^{*}$ (Corresponding Author) e-mail: aforaci@ufl.edu \\
$^{\dagger}$ e-mail: c.litos@ufl.edu \\
$^{\ddagger}$ e-mail: woodard@phys.ufl.edu \\
\end{flushleft}

\vskip 1cm 

\end{titlepage}

\newpage 

\section{Introduction}

The background geometry of cosmology is described in terms of its scale 
factor $a(t)$, its Hubble parameter $H(t)$, and its first slow roll 
parameter $\epsilon(t)$, 
\begin{equation}
    ds^2 = -dt^2 + a^2(t) d\vec{x} \cdot d\vec{x} \quad , \quad H(t) 
    \equiv \frac{\dot{a}}{a} \quad , \quad \epsilon(t) \equiv 
    -\frac{\dot{H}}{H^2} \; . \label{backg}
\end{equation}
The universe is said to be inflating when $H>0$ and $0\leq\epsilon<1$, 
and the data suggest this is happening now as well during the earliest 
instances of cosmic history \cite{Planck:2018vyg,AtacamaCosmologyTelescope:2025blo}. 
During an inflationary era virtual particles can be ripped out of the vacuum to 
become real \cite{Schrodinger}. Mass suppresses this effect, as does conformal
invariance \cite{Lifshitz:1945du}. Owing to their absence of mass and their lack of conformal invariance, 
massless, minimally coupled (MMC) scalars and gravitons 
\cite{Grishchuk:1974ny} experience the maximal effect, with the
occupation number of each wave vector growing like the square of the scale factor.
It is this staggering particle production which is believed to have caused the 
primordial tensor \cite{Starobinsky:1979ty} and scalar \cite{Mukhanov:1981xt} 
power spectra.

Inflationary particle production also has the consequence that correlators in 
theories involving MMC scalars and gravitons often inherit secular growth in 
their loop expansion, usually in powers of $\ln{[a(t)]}$. For example, on de
Sitter background ($\epsilon = 0$) the vacuum energy density and pressure of a 
MMC scalar with a $\frac{\lambda}{4!} \phi^4$ self interaction, where $\lambda$ is a dimensionless coupling constant, grow like 
\cite{Onemli:2002hr},
\begin{eqnarray}
\rho(t) & = \frac{3H^2}{8\pi G}+ \tfrac{\lambda H^4}{(4\pi)^4} \Bigl\{ 2 \ln^2(a) + 
\tfrac{13}{6} \ln(a) - \tfrac{43}{8} \Bigr\} + O(\lambda^2) \; , \label{rho} \\
p(t) & = -  \frac{3H^2}{8\pi G}+ \tfrac{\lambda H^4}{(4 \pi)^4} \Bigl\{-2 \ln^2(a) -
\tfrac{7}{2} \ln(a) + \tfrac{5}{3} \Bigr\} + O(\lambda^2) \; . \label{pressure}
\end{eqnarray}
Even when suppressed by the smallest coupling constants, the secular growth 
will eventually cause the breakdown of perturbation theory, necessitating 
some kind of resummation technique \cite{Woodard:2025smz}. The leading logarithm contributions 
in (\ref{rho}-\ref{pressure}) have two factors of $\ln(a)$ for each
additional $\lambda$ and are captured by Starobinsky's stochastic formalism
\cite{Starobinsky:1986fx,Tsamis:2005hd}. The same formalism can be resummed to 
give late time results \cite{Starobinsky:1994bd}.

Even though conformally invariant matter does not experience inflationary
particle production, a mismatch between the way primitive divergences and
counterterms depend on the scale factor still induces secular logarithms
\cite{Woodard:2025smz}. For example, a loop of photons on de Sitter 
background changes the Newtonian potential to \cite{Foraci:2024vng},
\begin{equation}
    \Psi = \tfrac{GM}{ar}\left\{1+\tfrac{4G}{15\pi a^2r^2} + 
    \tfrac{2GH^2}{5\pi^2}\ln(aHr)\; +\; \dots \right\} \;  .
\end{equation}
The 1-loop correction proportional to $r^{-2}$ is the de Sitter analogue of a 
well known correction on flat space \cite{Radkowski:1970ovx}, but the secular 
correction proportional to $\ln(aHr)$ is new, and its steady growth still
causes perturbation theory to break down. Because they derive from 
renormalization, it should not be surprising that these sorts of effects can 
be resummed using a variant of the renormalization group \cite{Woodard:2025smz}.

The late time behavior of matter loop corrections to gravity has been 
extensively studied for conformally invariant fields \cite{Foraci:2024vng,
Foraci:2024cwi} (massless conformally coupled scalars (MCC), fermions, photons),
and for MMC scalars \cite{Miao:2024atw}. The secular logarithms of conformal
matter can all be resummed using a variant of the renormalization group
\cite{Woodard:2025smz}. The same seemed to be true for MMC scalars
\cite{Miao:2024atw}, which was difficult to understand because MMC scalars
experience inflationary particle production and ought therefore to induce 
stochastic effects as well. The purpose of this paper is to show that they do.

Understanding the gravitational effects of MMC scalars has been a long and
confusing process. The initial computation of the graviton self-energy
\cite{Park:2011ww} was incomplete, and its use in solving the effective
field equation \cite{Park:2015kua} was therefore invalid, owing to a violation 
of the Ward identity which ultimately follows from stress-energy conservation 
\cite{Tsamis:2023fri}. This could be avoided by a finite renormalization of 
the cosmological constant, which is predicted by the stochastic formalism 
\cite{Miao:2024nsz}. Including this finite renormalization gave the right 
Newtonian potential \cite{Miao:2024atw}, but there was still an error in 
computing the effect on gravitational radiation, so that it still seemed as
if secular effects are entirely explained by a variant of the renormalization
group. By correcting this mistake we show that a loop of MMC scalars induces 
much stronger changes in gravitational radiation than those of a loop of 
conformal matter. This is the true effect of inflationary particle production
and we argue that it can be interpreted stochastically.

This paper is comprised of 5 sections, the first of which is this Introduction. Section 2 reviews how the 1-loop corrected graviton self-energy is used as a quantum source to correct the linearized field equations of gravity. Section 3 employs a Green's function method to obtain the exact 1-loop corrections to the graviton mode functions on de Sitter which are used to infer late time physics. Conclusions are reserved for section 4.  

\section{Corrections to the Graviton Self-Energy } \label{sec:Self-Energy} 
\begin{figure}[H]
          \centering
          \vskip 1cm
          \includegraphics[width=8cm]{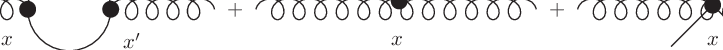}
          \caption{\footnotesize Diagrams which contribute to the 1-loop
          graviton self-energy. Curly lines represent gravitons and solid lines 
          represent an arbitrary matter field.}
         \label{diagrams}
         \end{figure}

Defining the graviton field $h_{\mu\nu}(x)$ for cosmology consists of subtracting away the background from the conformally rescaled metric tensor\footnote{We adopt a mostly positive Minkowski metic, $\eta_{\mu\nu}\equiv \text{diag}(-1,1,1,\dots,1) \; .$},
\begin{equation}
    g_{\mu\nu}(x)\equiv a^2[\eta_{\mu\nu}+\kappa h_{\mu\nu}(x)]\equiv a^2 \widetilde{g}_{\mu\nu}(x) \quad,\quad \kappa^2\equiv 16\pi G \; . \label{gravitonfield}
\end{equation}
The 1PI (one-particle-irreducible) 2-graviton function, denoted $-i[\mbox{}^{\mu\nu} \Sigma^{\rho\sigma}](x;x')$, is known as the graviton self-energy. When expressed in the Schwinger-Keldysh formalism, it can be used to correct the gravitational field equations by integrating it up against the graviton field and adding it as a quantum source in the linearized gravitational field equations \cite{Schwinger:1961Brownian,Mahanthappa:1962Photon,BakshiMahanthappa:1963I,BakshiMahanthappa:1963II,Keldysh:1965,Chou:1985,Jordan:1986,CalzettaHu:1987,FordWoodard:2005},
\begin{equation}
    \mathcal{L}^{\mu\nu\rho\sigma} \kappa h_{\rho\sigma}(x) - \int \!\! d^4x'
    \, [\mbox{}^{\mu\nu} \Sigma^{\rho\sigma}](x;x') \kappa h_{\rho\sigma}(x') 
    = \frac{\kappa^2}{2}T^{\mu\nu}(x) \; . \label{fullEOM}
\end{equation}
Here, $\mathcal{L}^{\mu\nu\rho\sigma}$ is the Lichnerowicz operator, and 
$T^{\mu\nu}(x)$ represents a classical source of stress-energy. This section is 
devoted to reviewing the 1-loop graviton self-energy due to loops of matter
fields, as shown in Figure~\ref{diagrams}.

The matter fields to be studied fall into two categories, the first of 
which consists of conformally invariant fields. This includes photons, 
massless Dirac fermions, and massless conformally coupled (MCC) scalars. 
Their respective Lagrangians are,
\begin{align}
    \mathcal{L}_{\text{EM}} & = -\tfrac14 F_{\mu\nu} F_{\rho\sigma} g^{\mu\rho}
    g^{\nu\sigma} \sqrt{-g}  \label{EM} \\
    \mathcal{L}_{\text{MCC}} &= -\tfrac12 \partial_\mu \phi \partial_\nu \phi 
    g^{\mu\nu} \sqrt{-g} - \tfrac18 \tfrac{(D-2)}{(D-1)} R \phi^2 \sqrt{-g} 
    \label{MCC} \\
    \mathcal{L}_{\text{Dirac}} & = \overline{\psi} e^{\mu}_{~a} \gamma^a \Bigl(
    i \partial_{\mu} \!-\! \tfrac12 A_{\mu bc} J^{bc} \Bigr) \psi \sqrt{-g} 
    \label{Dirac} \; ,
\end{align} 
where $A_{\mu bc}$ is the spin connection, ${e^{\mu}}_{a}$ is the vierbein field, 
and $J^{bc}$ are the generators of the bispinor representation of the Lorentz 
group in \cref{Dirac}. Note that electromagnetism (\ref{EM}) is conformally 
invariant in $D=4$, while the latter two theories are conformally invariant for 
arbitrary $D$. Regardless of the matter loop that appears, at most two BPHZ 
counterterms (Bogoliubov, Parasiuk \cite{Bogoliubov:1957gp,Hepp:1966eg,
Zimmermann:1968mu,Zimmermann:1969jj}) are required to subtract away the 
1-loop divergences of the theory,
\begin{equation}
\Delta\mathcal{L} = c_1 R^2 \sqrt{-g} + c_2 C^{\alpha\beta\gamma\delta}
C_{\alpha\beta\gamma\delta} \sqrt{-g} \; .
\end{equation}

When the diagrams in Figure~\ref{diagrams} are computed for conformally
invariant matter (\ref{EM}-\ref{Dirac}), using dimensional regularization, 
and BPHZ renormalization, the result is,
\begin{equation}
    [\mbox{}^{\mu\nu} \Sigma_{\rm}^{\rho\sigma}](x;x') = -
\tfrac{\kappa^2}{q \pi^3} \, 
\mathcal{C}^{\alpha\beta\gamma\delta\mu\nu} \!\times\!
{\mathcal{C}'}_{\alpha\beta\gamma\delta}^{~~~~~\rho\sigma} \Bigl[ 8 \pi 
\ln(a a') \delta^4(x \!-\! x') + f_B(x \!-\! x')\Bigr] \; , \label{confSelfEnergy}
\end{equation}
where $q$ depends on the field involved in the loop correction,
\begin{equation}
     q = \begin{cases}
        2^{10} \cdot 3 \cdot 5 \,, \ \ \text{MCC scalars} \\   
        2^9 \cdot 5 \,, \qquad \text{fermions} \\
        2^8 \cdot 5 \,, \qquad \text{photons}
    \end{cases} \label{qdef}
\end{equation}
The function $f_B(x-x')$ is defined as,
\begin{equation}
f_B(x \!-\! x') \equiv \partial^4 \Bigl\{ \theta(\Delta \eta \!-\! \Delta r)
\Bigl(\ln[\mu^2 (\Delta \eta^2 \!-\! \Delta r^2)] \!-\! 1 \Bigr) \Bigr\} 
\quad , \quad \Delta r\equiv \lVert\vec{x}-\vec{x}'\rVert  \; , \label{fB}
\end{equation}
and the tensor differential operator $\mathcal{C}_{\alpha\beta\gamma\delta}^{
~~~~~\mu\nu}$ can be read off from the linearized, conformally rescaled Weyl tensor, 
\begin{equation}
    \widetilde{C}_{\alpha\beta\gamma\delta} = {\mathcal{C}_{\alpha\beta\gamma\delta}}^{
    \mu\nu}\times \kappa h_{\mu\nu} + O(h^2) \; . \label{cOperator}
\end{equation}
It is significant to note that the conformal invariance of MCC scalars and fermions 
renders their primitive contributions to (\ref{confSelfEnergy}) identical to their 
flat space counterparts \cite{Duff:2000mt}. This means that the scale factor enters 
the calculation only through renormalization in which no assumptions about its 
explicit form are made, thus (\ref{confSelfEnergy}) is valid on any cosmological 
background for these two fields.

The second category consists of fields which are not conformally invariant, and 
the only self-energy worked out to date is due to a loop of MMC scalars,
\begin{equation}
    \mathcal{L}_{\text{MMC}} = -\tfrac12\partial_\mu\phi\partial_\nu\phi g^{\mu\nu}\sqrt{-g} \; .
\end{equation}
It's contribution to the graviton self-energy is a complicated expression that can be found explicitly in \cite{Miao:2024atw,Miao:2024pwd}. One feature of the graviton self-energy 
due to a loop of MMC scalars is that its primitive contribution is not conserved. This is circumvented by adding to the theory a finite cosmological counterterm \cite{Tsamis:2023fri},
\begin{equation}
    \Delta\mathcal{L}_\Lambda = \gamma\sqrt{-g} \; .
\label{eq:countertermlambda}\end{equation}
The coefficient $\gamma$ is finite in $D=4$,
\begin{equation}
    \gamma \rightarrow -\frac{H^4}{8\pi^2}\; ,
\label{eq:gammavalue}\end{equation}
and this ultimately induces a fractional change in the cosmological constant. Finite renormalizations are usually optional, but in this theory it is required to enforce conservation.

\section{Correction to the Mode Functions} 

By setting the stress tensor to zero one can study the effects of loops of fields on the plane wave gravitons, 
\begin{equation}
    \kappa h_{\mu\nu} = \e_{\mu\nu} e^{i\vec{k} \cdot \vec{x}} u(t,k) \;,
\label{eq:planewaves}\end{equation}
where the polarization tensor $\e_{\mu\nu}$ takes the same form as in flat space and $u(t,k)$ is the mode function. In what follows, we shall study the secular effects induced by fields on a background de Sitter cosmology. In that case, for Bunch-Davies initial conditions the tree order result takes the following form, 
\begin{equation}
    u^{\rm tree}(t,k) = \frac{H}{\sqrt{2k^3}} \left[1 - \frac{ik}{aH}\right] \exp\left[\frac{ik}{aH}\right] \; .
\label{eq:BD}\end{equation}
On de Sitter background, the Lichnerowicz operator reduces to, 
\begin{equation}
\begin{aligned}
    \mathcal{L}^{\mu\nu\rho\sigma} \kappa h_{\rho \sigma} &= \frac{a^2}{2} \left[ \p^2 h^{\mu\nu} - \eta^{\mu\nu} \p^2 h + \eta^{\mu\nu} \p^\rho \p^\sigma h_{\rho\sigma} + \p^\mu \p^\nu h - 2 \p^\rho \p^{(\mu} h^{\nu)}_{\ \rho}\right]\\ 
    &+a^3 H \left[ \eta^{\mu\nu} \p_0 h -\p_0 h^{\mu\nu} - 2 \eta^{\mu\nu} \p^\rho h_{\rho0} + 2 \p^{(\mu}h^{\nu)}_{\ 0}\right] + 3a^4 H^2 \eta^{\mu\nu} h_{00} \; .
    \end{aligned}
\label{eq:Lichn}\end{equation}

The general form of the equation for the 1-loop correction graviton mode function 
takes the form,
\begin{equation}
    \left[ \p_t^2 + 3H \p_t + \frac{k^2}{a^2}\right] \kappa^2 u_1 = S(t) \, \, , \, \, u(t,k) = u^{\rm tree}(t,k) + \kappa^2 u_1(t,k) + \mathcal{O}(\kappa^2) \; .
\label{eq:EOM}\end{equation}
Here $\kappa^2 u_1(t,k)$ is the 1-loop correction to the mode function and 
$S(t)$ is a "source" function depending on the field under consideration. The 
previous equation can be formally solved via the Greens function method,
\begin{equation}
    \kappa^2 u_1(t,k) = \!\!\!\int_{0}^\infty dt'\ G(t;t') S(t') \; .
\label{eq:GFmethod}\end{equation}
where the (retarded) Greens function can be written in terms of the 
tree-order mode function, 
\begin{equation}
\begin{aligned}
    G(t;t') &= i a^3(t') \theta(t-t') \left[ u^{\rm tree}(t) u^{\ast{\rm tree}}(t') - u^{\ast{\rm tree}}(t) u^{\rm tree}(t')\right]\; .
\end{aligned}
\label{eq:GreensFunction}\end{equation}
This allows the 1-loop correction to the mode function to be expressed as a linear 
combination of the tree-order mode function and its conjugate, 
\begin{equation}
    \kappa^2 u_1(t,k) = \alpha(t,k) u^{\rm tree}(t,k) + \beta(t,k) u^{\ast {\rm tree}}(t,k) \; ,
\label{eq:u1gensolution}\end{equation}
with the functions $\alpha$ and $\beta$ being defined as, 
\begin{equation}
    \alpha(t,k) \equiv i \int_0^t dt'\ a^3(t') u^{\ast {\rm tree}}(t',k) S(t') \, \, , \, \, \beta(t,k) \equiv -i \int_0^t dt'\ a^3(t') u^{{\rm tree}}(t',k) S(t') \; .
\label{eq:coefficientsalphabeta}\end{equation}

It is well known that inflationary perturbations generate power spectra which
are approximately scale-invariant \cite{Weinberg:2008zzc}. The short-wavelength $k$-modes cross the 
horizon during inflation, become super-horizon, and then freeze in to a 
constant amplitude. This is easy to see for the tree order mode function
(\ref{eq:BD}). The maximum growth that loop corrections can experience is
bounded by powers of $\ln(a)$ due to an important theorem of Weinberg 
\cite{Weinberg:2006ac}. It turns out that single matter loop corrections to
$u(t,k)$ induce no growth, but they can change asymptotic freeze-in value,
as well as the rate at which freeze-in occurs. The general late time form is,
\begin{equation}
    \kappa^2 u_1 = \frac{\kappa^2 H^2}{\pi^2} \times \frac{H}{\sqrt{2k^3}} \times\left[ \mathcal{A} + \mathcal{B} \times \left( \frac{k}{aH} \right)^2 + i\mathcal{C} \times \ln(a) \times \left( \frac{k}{aH} \right)^3 +  \mathcal{O}\left(\frac{k}{aH}\right)^5  \right] \; .
\label{eq:latetimelimitmodefunction}\end{equation} 
The coefficients $\mathcal{A}, \mathcal{B}$ and $\mathcal{C}$ depend on the field 
under consideration and will be derived in the following subsections. For now, 
let us note that the "electric" part of the Weyl tensor is\cite{Miao:2024atw}, 
\begin{equation}
\begin{aligned}
C_{0i0j} &= \e_{ij} e^{i\vec{k} \cdot \vec{x}}\times -\frac{1}{4}\left( \p_0^2 - k^2 \right) u(t,k)\\ 
&=-\frac{i\sqrt{2k^3}}{4a} \e_{ij} e^{i\vec{k} \cdot \vec{x}} \times \left\{1 - \frac{i\kappa^2 H^2}{\pi^2} \left( \mathcal{B} - \frac{\mathcal{A}}{2}\right) \left(\frac{aH}{k}\right)+ \frac{3 \kappa^2 H^2}{\pi^2} \mathcal{C} \ln(a) +\ldots\right\}\\ 
&\equiv C_{0i0j}^{\rm tree} \times  \left\{1 - \frac{i\kappa^2 H^2}{\pi^2} \left( \mathcal{B} - \frac{\mathcal{A}}{2}\right) \left(\frac{aH}{k}\right)+ \frac{3 \kappa^2 H^2}{\pi^2} \mathcal{C} \ln(a) +\ldots\right\}
\end{aligned}
\label{eq:electricWeyltensor}\end{equation}
The correction proportional to $\mathcal{C}$ arises from renormalization 
\cite{Woodard:2025smz}, as we have discussed. When $\mathcal{B} \neq \frac12 
\mathcal{A}$ there is a much stronger effect. We will show that this is
absent for conformally invariant matter but that it occurs for MMC scalars.  

\subsection{Corrections from Conformally Coupled Fields}

Since the graviton self-energy resulting from a loop of photons, fermions, and MCC scalars all take the same form up to the field-dependent parameter $q$ defined in (\ref{qdef}), they share a quantum source of the same form in the effective field equations. The source that gives rise to the 
corrections is,
\begin{equation}
    S_q = \frac{32 \kappa^2 H^2}{q \pi^2} \cdot -\frac{ik^3}{H a^3} \cdot
    \exp \left[\frac{ik}{aH}\right] \, \, . \, \, 
\label{eq:CCsource}\end{equation}
The resulting coefficient functions are found via substitution of (\ref{eq:CCsource}) into (\ref{eq:coefficientsalphabeta}), 
\begin{align}
    \alpha(t,k) &= \frac{16  \kappa^2 H^2}{q\pi^2} \left[\ln(a) - \frac{ik}{aH}(1-a)\right],\\ 
    \beta(t,k) &= \frac{16 \kappa^2 H^2}{q\pi^2} \left[\frac{1}{2} \left( e^{\frac{2ik}{H}} - e^{\frac{2ik}{aH}} \right)+ {\rm Ei}\left[\frac{2ik}{aH}\right] - {\rm Ei}\left[\frac{2ik}{H}\right]  \right]. 
\end{align}
We read off the following values for the constants $\mathcal{A},\mathcal{B}$ 
and $\mathcal{C}$:
\begin{equation}
    \mathcal{A}_{q} = \frac{8}{q} \left[ 2\gamma-1+e^{\frac{2ik}{H}}+\tfrac{2ik}{H}+i\pi-2\operatorname{Ei}(\tfrac{2ik}{H})+2\ln{(\tfrac{2k}{H})}\right] \, , \, \mathcal{B}_q = \frac{\mathcal{A}_q}{2} \, , \, \mathcal{C}_q = \frac{32}{3q} \; .
\label{eq:ABCq}
\end{equation}
The relation between $\mathcal{A}_q$ and $\mathcal{B}_q$ demonstrates that the correction in \cref{eq:electricWeyltensor} vanishes. The secular growth implied by $\mathcal{C}_{q}$
has been studied before and can be summed up using a variant of the renormalization
group \cite{Foraci:2024cwi,Woodard:2025smz}.

\subsection{Corrections from Massless, Minimally Coupled Scalars}

MMC scalars differ from conformal matter because the source decays more slowly,
\begin{equation}
S_{\rm MMC}(t,k) = \frac{\kappa^2 H^2}{48\pi^2} \cdot \frac{k^2}{a^2} 
\cdot \frac{H}{\sqrt{2k^3}} \cdot \left(1 - \frac{ik}{10aH} \right) \exp 
\left[\frac{ik}{aH}\right] \; .
\label{eq:MMCsource}\end{equation}
The associated coefficient functions $\alpha$ and $\beta$ are,
\begin{align}
    \alpha(t,k) &= \frac{\kappa^2 H^2}{96\pi^2} \left[i \left(1 - a^{-1}\right)\left( \frac{aH}{k} + \frac{k}{10H}\right) - \frac{9}{10} \ln(a) \right] \; ,\\ 
    \beta(t,k) &= \frac{\kappa^2 H^2}{96\pi^2} \left[\frac{e^{\frac{2ik}{H}} - e^{\frac{2ik}{aH}}}{20} + \frac{iaH}{k} \left(\frac{1}{a} e^{\frac{2ik}{H}}-e^{\frac{2ik}{aH}} \right)+ \frac{9}{10} \left( {\rm Ei}\left[\frac{2ik}{H}\right]-{\rm Ei}\left[\frac{2ik}{aH}\right] \right)\right] \; .
\end{align}
Taking the late time limit gives,
\begin{equation}
    \mathcal{A}_{\rm MMC} = \frac{1}{96}\left(\frac{ik}{10H} +\left(e^{\frac{2ik}{H}}-1\right)\left(\frac{iH}{k} +\frac{1}{20}\right)- \frac{9}{10} \left[ \gamma + \frac{i\pi}{2} + \ln\left[\frac{2k}{H}\right] -{\rm Ei}\left[\frac{2ik}{H}\right]\right]\right) \; ,
\label{eq:AMMC}\end{equation}
\begin{equation}
\mathcal{B}_{\rm MMC} =  \left[ \frac{\mathcal{A}_{\rm MMC}}{2} -\frac{1}{96}\right] \quad,  \quad \mathcal{C}_{\rm MMC} = - \frac{1}{160} \; .
\label{eq:BCMMC}\end{equation}
We find the late time Weyl tensor from expression (\ref{eq:electricWeyltensor}),
\begin{equation}
    C_{0i0j} = C_{0i0j}^{\rm tree} \times \left\{ 1 + \frac{i\kappa^2 H^2}{96\pi^2} 
    \frac{aH}{k}  - \frac{3\kappa^2 H^2}{160\pi^2} \ln(a) + \ldots \right\}
    \label{eq:electricWeyltensorMMC}
\end{equation}

A few comments are in order. First, the logarithmic term appearing in
\cref{eq:electricWeyltensorMMC} would correspond to $q = -2^{10} \cdot 5/3$ 
were this a conformally coupled field. Second, the much stronger effect from
$\mathcal{B} \neq \frac12 \mathcal{A}$ arises from the copious production of
MMC scalars during inflation. These particles cannot actually induce any growth
in the graviton mode function because they only couple to gravity through their
rapidly red-shifting kinetic energies, but they do reduce the rate at which the
graviton mode function freezes in. The fact that this effect cannot be captured
using the renormalization group suggests that we seek a stochastic origin for 
it. That is the subject of the next subsection.

\subsection{Shifting the cosmological constant}

Consider making two changes on the tree order mode function:
\begin{enumerate}
\item{Rescale the entire result by a constant field strength $\sqrt{Z} = 
1 + \frac12 \delta Z$; and}
\item{Rescale the Hubble parameter $H \rightarrow H + \delta H$.}
\end{enumerate}
The resulting change in the late time form is,
\begin{equation}
\delta u^{\rm tree} \longrightarrow \tfrac12 \delta Z \times 
\tfrac{H}{\sqrt{2 k^3}} \Bigl\{1 + \tfrac12 (\tfrac{k}{a H})^2 + \dots \Bigr\} 
+ \tfrac{\delta H}{H} \times \tfrac{H}{\sqrt{2 k^3}} \Bigl\{ 1 - \tfrac12
(\tfrac{k}{a H})^2 + \dots \Bigr\} \; . \label{rescale}
\end{equation}
Comparison with the asymptotic form (\ref{eq:latetimelimitmodefunction})
allows us to determine $\delta Z$ and $\delta H/H$,
\begin{equation}
\left\{ \begin{matrix}
\tfrac{\kappa^2 H^2}{\pi^2} \mathcal{A} = \tfrac12 \delta Z + \tfrac{\delta H}{H} \\
\tfrac{\kappa^2 H^2}{\pi^2} \mathcal{B} = \tfrac14 \delta Z - \tfrac12 \tfrac{\delta H}{H} 
\end{matrix} \right\}
\qquad \Longrightarrow \qquad
\left\{ \begin{matrix}
\delta Z = \tfrac{\kappa^2 H^2}{\pi^2} (\mathcal{A} + 2\mathcal{B} ) \\
\tfrac{\delta H}{H} = \tfrac{\kappa^2 H^2}{\pi^2} (\tfrac12
\mathcal{A}-\mathcal{B})  
\end{matrix} \right\} \; . \label{deltaZH}
\end{equation} 
Because conformally invariant matter loops obey $\mathcal{B} = \frac12 \mathcal{A}$,
it follows that they induce no change in the Hubble parameter. However, the
MMC scalar result (\ref{eq:BCMMC}) implies a 1-loop change in the Hubble
parameter of,
\begin{equation}
\delta H =  \tfrac{\kappa^2 H^3}{96\pi^2} \; .
\label{eq:deltaHfinalresult}\end{equation}

On de Sitter, the shift (\ref{eq:deltaHfinalresult}) of the Hubble parameter 
is tantamount to a shift of the cosmological constant, 
\begin{equation}
    \delta \Lambda = 6 H\delta H =  \tfrac{\kappa^2 H^4}{16\pi^2} \; .
\label{eq:shiftinLambda}\end{equation}
This could be accomplished by the finite renormalization,
\begin{equation}
\Delta \mathcal{L}_{\rm shift} = -\tfrac{2}{\kappa^2} \delta \Lambda \sqrt{-g} 
= -\tfrac{H^4}{8\pi^2} \sqrt{-g} = \gamma \sqrt{-g} \; , 
\end{equation}
where $\gamma$ was defined in \cref{eq:gammavalue}. This happens to be exactly
the finite renormalization required for conservation \cite{Tsamis:2023fri} and
predicted by the stochastic formalism \cite{Miao:2024nsz}. 

Note also that shifting the Hubble parameter (\ref{eq:deltaHfinalresult})
induces no change in the 1-loop correction to the Newtonian potential because
the tree order Newtonian potential ($\Psi = GM/ar$) does not depend on $H$. 
This stochastic explanation therefore explains both the large correction (\ref{eq:electricWeyltensorMMC}) to the Weyl tensor of gravitational radiation 
and the fact that the Newtonian potential induced by MMC scalars experiences
only the renormalization group effect. 

\section{Epilogue}

The study of quantum gravity on de Sitter background has been a long 
and arduous endeavor. Of particular interest are the secular effects
that tend to occur in loop corrections. Recent work suggests that the 
leading logarithm corrections can be captured by a combining a variant
of Starobinsky's stochastic formalism with a variant of the renormalization 
group \cite{Woodard:2025smz}. Matter loop corrections provide an especially
simple venue because they are gauge independent. Conformally invariant
matter does not experience inflationary particle production, so the only
secular factors from loops of conformal matter derive from the mismatch
between how primitive divergences and counterterms depend on the scale
factor \cite{Woodard:2014jba}. In contrast, MMC scalars experience explosive particle production \cite{Woodard:2014jba}, 
so the secular growth they induce sometimes has a stochastic explanation 
in addition to the universal renormalization effect \cite{Miao:2024nsz}. We have argued that
this occurs for MMC scalar loop corrections to gravitational radiation.

Expression (\ref{eq:latetimelimitmodefunction}) gives the late time form
of matter loop corrections to the mode function of gravitational radiation
in terms of three constants $\mathcal{A}$, $\mathcal{B}$ and $\mathcal{C}$.
The constant $\mathcal{C}$ derives from the universal renormalization
effect. The constants $\mathcal{A}$ and $\mathcal{B}$ can be resolved into
changes to the field strength and to the Hubble parameter (\ref{deltaZH}).
Conformally invariant matter induces no change in the Hubble parameter,
but MMC scalars cause a shift (\ref{eq:deltaHfinalresult}) which is
predicted by the stochastic formalism. The physical origin of this shift
is the inflationary production of MMC scalars. These quanta induce no
growth of the mode function itself because they couple to gravity only 
through their rapidly red-shifting kinetic energies, however, they do
change the rate at which the mode function freezes in.

Two points require additional work. The first of these concerns the 1-loop
corrections to the scalar field strength which can be read off from the
values of $\mathcal{A}$ and $\mathcal{B}$ through expression (\ref{deltaZH}).
Of course these would multiply the tensor power spectrum. They are suppressed 
by the small loop-counting parameter $\kappa^2 H^2 \ltwid 10^{-10}$ but their
complicated dependence on $k/H$ might give rise to observable effects. A key
issue is whether or not $\delta Z$ can be absorbed into perturbative 
corrections to the initial state of the sort already studied for scalar
potential models \cite{Kahya:2009sz}. 

The second point is whether or not graviton loop corrections can induce
stronger effects. Weinberg's theorem \cite{Weinberg:2006ac} actually allows 
for logarithmic growth of the mode function. This is not induced by MMC 
scalar loops because MMC scalar couple to gravity only through their rapidly
red-shifting kinetic energies. However, gravitons are also produced copiously
during inflation, and they possess a spin-spin coupling which does not 
red-shift \cite{Tan:2021lza}. A plausible extension of an unregulated result for the graviton 
self-energy \cite{Tsamis:1996qk} suggests secular growth of the mode function 
\cite{Tan:2021lza}. It would be fascinating to see what a dimensionally 
regulated and fully renormalized result gives.

\vskip 1cm

\centerline{\bf Acknowledgements}

This work was partially supported by NSF grant PHY-2207514 and by
the Institute for Fundamental Theory at the University of Florida.

\end{document}